\documentclass[sigconf]{acmart}

\renewcommand\footnotetextcopyrightpermission[1]{}
\AtBeginDocument{%
  \providecommand\BibTeX{{%
    \normalfont B\kern-0.5em{\scshape i\kern-0.25em b}\kern-0.8em\TeX}}}
    
\usepackage{lipsum}
\usepackage{enumitem}
\usepackage[framemethod=TikZ]{mdframed}
\usepackage{soul}
\usepackage{listings}
\usepackage{xspace}
\usepackage{hyperref}

\usepackage[font=small,labelfont=bf]{caption} %
\setlist{noitemsep}

\usepackage{siunitx}
\def\eg{\emph{e.g.,}\xspace}

\def\ie{\emph{i.e.,}\xspace}
\def\etal{\emph{et al.}\xspace}

\newtheorem{definition}{Definition}

\newcommand{\adv}[0]{\texttt{Adv}\xspace}
\newcommand{\dr}[0]{\texttt{Det}\xspace}
\newcommand{\imp}[0]{\texttt{Imp}\xspace}

\newcommand{\uf}[0]{\texttt{UI\_features}\xspace}
\newcommand{\pk}[0]{\texttt{Pre-knowledge}\xspace}
\newcommand{\se}[0]{\texttt{Sequence}\xspace}

\newcommand{\low}[0]{\texttt{low}\xspace}
\newcommand{\medium}[0]{\texttt{medium}\xspace}
\newcommand{\high}[0]{\texttt{high}\xspace}

\newcommand{\uiguard}[0]{\textsc{UIGuard}\xspace}
\newcommand{\aidui}[0]{\textsc{AidUI}\xspace}

\newcommand{\PZwithC}[0]{\num{1.4}\xspace}
\newcommand{\PZwithoutC}[0]{\num{2.7}\xspace}
\newcommand{\advPZ}[0]{\num{0.22}\xspace}
\newcommand{\fscorePZ}[0]{\num{0.73}\xspace}

\newcommand{\PRwithC}[0]{\num{3.1}\xspace}
\newcommand{\PRwithoutC}[0]{\num{4.9}\xspace}
\newcommand{\advPR}[0]{\num{0.1}\xspace}
\newcommand{\fscorePR}[0]{\num{0.18}\xspace}

\newcommand{\PAwithC}[0]{\num{3.7}\xspace}
\newcommand{\PAwithoutC}[0]{\num{2.3}\xspace}
\newcommand{\advPA}[0]{\num{0.82}\xspace}
\newcommand{\fscorePA}[0]{\num{0.77}\xspace}

\newcommand{\RMwithC}[0]{\num{8.3}\xspace}
\newcommand{\RMwithoutC}[0]{\num{6.1}\xspace}
\newcommand{\advRM}[0]{\num{0.82}\xspace}
\newcommand{\fscoreRM}[0]{\num{0.18}\xspace}

\begin{document}

\title{The Invisible Game on the Internet: \\A Case Study of Decoding Deceptive Patterns}

\begin{abstract}
Deceptive patterns are design practices embedded in digital platforms to manipulate users, representing a widespread and long-standing issue in the web and mobile software development industry. 
Legislative actions highlight the urgency of globally regulating deceptive patterns. However, despite advancements in detection tools, a significant gap exists in assessing deceptive pattern risks.
In this study, we introduce a comprehensive approach involving the interactions between the Adversary, Watchdog (e.g., detection tools), and Challengers (e.g., users) to formalize and decode deceptive pattern threats. Based on this, we propose a quantitative risk assessment system. Representative cases are analyzed to showcase the practicability of the proposed risk scoring system, emphasizing the importance of involving human factors in deceptive pattern risk assessment.
\end{abstract}

\author{
Zewei Shi, Ruoxi Sun, Jieshan Chen, Jiamou Sun, and Minhui Xue}
\affiliation{
\institution{CSIRO's Data61, Australia}\country{}
}

\maketitle

\pagestyle{plain}

\section{Introduction}

Deceptive design practices have emerged as a pervasive challenge in contemporary digital landscapes, exerting a profound influence on user experiences across diverse platforms such as social media, e-commerce, mobile devices, cookie consent banners, and gaming. These manipulative strategies, often referred to as dark patterns, are purposefully embedded within the user interfaces (UIs) of websites and applications, strategically employed by companies to extract profit, harvest data, and curtail consumer choice~\cite{gray2018dark, mathur2019dark,di2020ui}. They encompass a spectrum of deceitful techniques, including the use of exaggerated language, social proof featuring fake or selective endorsements, incessant nagging with misleading information, and the propagation of alternative facts~\cite{cprc2022dark}.

In 2022, updates to the EU’s Digital Services Act (DSA) included a ban on ``dark patterns'' designed to deceive or manipulate users~\cite{ep2022digital}. Beyond the EU, the California Consumer Privacy Rights Act (CPRA) has specifically called out dark patterns in the context of valid user consent to data processing~\cite{gole2022new}. A recent survey in Australia indicates that 83\% of Australians encountered adverse consequences due to design features on digital platforms~\cite{cprc2022dark}. 
France’s data protection agency has fined Google €150 million and Facebook €60 million for making it too confusing for users to reject cookies~\cite{vincent2022france}. 
Similarly, the White House recently reacted to the explicit, AI-generated images of music superstar Taylor Swift that had gone viral online, calling it ``alarming'', and leaned on Congress for a legislative crackdown~\cite{rahman2024taylor}. 
In the context of national defense, deceptive patterns could lead to phishing attacks or social engineering, where misleading interfaces trick military personnel into revealing sensitive information, spreading disinformation, manipulating public opinion on defense, or influencing political stances. Notably, Forbes has recognized the role of social media (\eg X) in disseminating disinformation, particularly in the context of the Russia-Ukraine war~\cite{suciu2023x}.  

Recent research efforts have made significant strides in proposing taxonomies to categorize the diverse types of deceptive patterns~\cite{gray2023towards, gray2018dark, chen2023unveiling, di2020ui}. Additionally, various researchers have engaged in distinct systematic empirical studies to comprehend and evaluate the prevalence of deceptive patterns across different digital platforms~\cite{di2020ui,gray2018dark}.
From the perspective of deceptive pattern detection, Mathur~\etal~\cite{mathur2019dark} identified 1,818 instances of deceptive commercial patterns among approximately 11,000 websites affiliated with retail businesses and online marketplaces.
Recently, Chen~\etal have developed a knowledge-driven system, named \uiguard{}~\cite{chen2023unveiling}. This innovative tool employs computer vision and natural language pattern matching to autonomously identify a wide array of deceptive patterns in mobile user interfaces.

Despite these advancements in detection tools, there still exists a significant gap in the field of deceptive pattern risk assessment, while publicly accessible issue disclosing and tracking websites are widely developed in security domain, such as the well-known CVE.\footnote{\url{https://www.cve.org/}} Further research and development are needed to bridge this gap and enhance our understanding of the risks posed by deceptive user interface design practices.
In this study, we aim to formalize and decode the threats of deceptive patterns and propose a practical risk assessment approach to systematically evaluate and analyze deceptive pattern implementations in the real world.
The key contribution of this study is threefold:

\begin{itemize}[leftmargin=*]
\item We propose a new approach to decode and understand the threats introduced by deceptive patterns within a security game context, which innovatively involves the Adversary, Watchdog (\eg detection tools), and Challengers (\eg users) to build up a comprehensive threat model. 
\item Based on the deceptive pattern game, we next introduce a quantitative approach to assess the risk of a specific deceptive pattern implementation, taking the interactions between the adversary, watchdog, and challengers into account while incorporating the impact of a deceptive pattern's consequence as well.
\item We applied the proposed risk scoring system to various deceptive pattern categories. To demonstrate the practicability of the system, we report several representative cases with detailed analyses of their risk levels and potential consequences. We further showcased the necessity of involving the human factors in deceptive pattern risk assessment.
\end{itemize}

\section{Decoding Deceptive Pattern with Game-based Security}

In this section, we present a new threat modeling approach designed to analyze deceptive patterns within a gaming context, encompassing perspectives from attackers, watchdogs, and challengers. We then introduce a scoring system that offers a quantitative assessment of the risk associated with a particular deceptive pattern implementation. 

\subsection{Deceptive Pattern Game}\label{sec_dark_pattern_game}

Different from other existing studies that focus on the identification and detection of deceptive patterns in software implementations, inspired by a cryptographic game introduced by Cliptography~\cite{russell2016cliptography}, in this study, we introduce a security game to formulate deceptive patterns. Concretely, a software scheme $\Pi$ consists of a set of functionalities ($F^1, F^2, ... , F^k$). 
The definition of $\Pi$ results in a specification of the software; for concreteness, we label these as $\Pi_{s} = (F_s^1, F_s^2, ... , F_s^k)$; when a scheme is (perhaps maliciously) implemented, we denote the implementation as $\Pi_{i} = (F_i^1, F_i^2, ... , F_i^k)$. If the implementation honestly follows the specification of the scheme, we have $\Pi_s = \Pi_i$.

In our definition, the adversary $\mathcal{A}$ will interact with both the watchdog $\mathcal{W}$ (\ie deceptive pattern detection measures, such as app store policies, regulations, and security evaluation tools that provide protection) and the challenger $\mathcal{C}$ (\ie human perspective, such as users' common sense and behavior, that helps users avoid being misled by deceptive patterns). The adversary provides $\mathcal{W}$ his potentially subverted implementations $\Pi_i$ of the primitive (as oracles); $\mathcal{W}$ may then interrogate them in an attempt to detect divergence from the specification. On the basis of these tests, the watchdog produces a bit, indicating whether $\mathcal{A}$ wins the game (\ie the implementations passed whatever tests the watchdog carried out to detect inconsistencies with the specification). 

\begin{definition}
A deceptive pattern game $G = (\mathcal{C},\Pi_s)$ is defined by a challenger $\mathcal{C}$ and a specification $\Pi_s$. Given an adversary $\mathcal{A}$ and a watchdog $\mathcal{W}$, we define the \textbf{detection probability} of the watchdog $\mathcal{W}$ with respect to $\mathcal{A}$ to be
\begin{equation}
\dr_{\mathcal{W},\mathcal{A}} = |Pr[\mathcal{W}(\Pi_{i}) = 1] - Pr[\mathcal{W}(\Pi_{s}) = 1]|,
\end{equation}
where $\Pi_{i} = (F_i^1, F_i^2, ... , F_i
^k)$ denotes the implementation produced by $\mathcal{A}$ and $\Pi_{s}$ represents the specification. The \textbf{advantage of the adversary} is defined to be
\begin{equation}
\adv_{\mathcal{A},\mathcal{C}} = Pr[(\mathcal{A} \Leftrightarrow \mathcal{C}(\Pi_{i})) = 1],   
\end{equation}
where $\mathcal{A} \Leftrightarrow \mathcal{C}(\Pi_{i})$ denotes the interaction between $\mathcal{A}$ and $\mathcal{C}$, which returns 1 when $\mathcal{A}$ wins the game.
We say that a game is deceptive pattern-resistant if there exists a watchdog $\mathcal{W}$ such that for an adversary $\mathcal{A}$, either $\dr_{\mathcal{W},\mathcal{A}}$ is non-negligible, or $\adv_{\mathcal{A},\mathcal{C}}$ is negligible.
\end{definition}

\begin{figure}[t]
\centering
\includegraphics[width=0.85\linewidth]{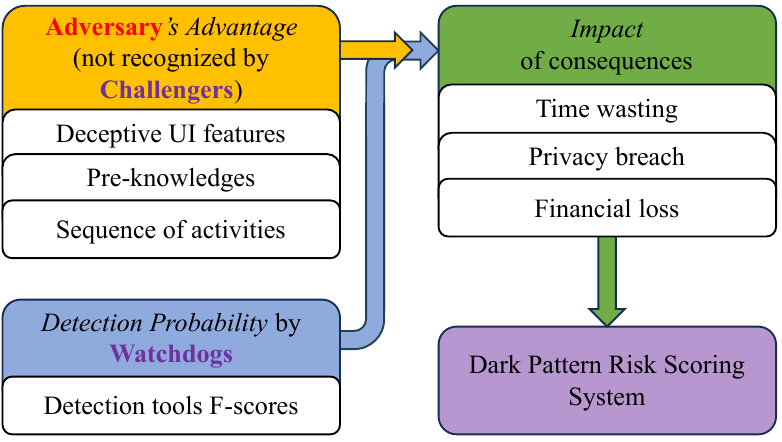}
\caption{An Overview of Deceptive Pattern Risk Scoring System.}
\label{fig_overview}
\end{figure}

\subsection{Deceptive Pattern Risk Scoring System}

Based on the modeled deceptive pattern game, we further propose an assessment approach to evaluate the risk of a specific software implementation, especially focusing on user interfaces. The risk scoring system considers both adversarial and watchdog perspectives with regard to the detection probability (\dr) and the adversary's advantage (\adv). Our deceptive pattern game particularly considers human factors in the definition of \adv.
Additionally, we incorporate the impact of a deceptive pattern's consequences as a factor in the scoring system, as illustrated in Figure~\ref{fig_overview}.

As previously defined, the detection probability (\dr) refers to the capability of a watchdog to detect deceptive patterns in a software implementation while considering its specification. In practice, such a capability could be represented through the F-score of a detection tool, since it involves both precision (true deceptive patterns in positive predictions, reflecting $Pr[\mathcal{W}(\Pi_{i}) = 1]$) and recall (true deceptive patterns in positive ground truths, related to $Pr[\mathcal{W}(\Pi_{s}) = 1]$) simultaneously. In our study, we derive \dr from the F-scores of \uiguard{}~\cite{chen2023unveiling}, a state-of-the-art deceptive pattern detection tools.
To determine the adversary's advantage (\adv), we initially consider three sub-factors: \uf, \pk, and \se. Specifically, \uf refers to whether the user interface contains low/medium/high-risky deceptive features that could mislead or fool a user, such as small icon, double negation, and inconsistent UI metaphor; \pk considers whether a user needs a good enough background or professional knowledge to recognize a deceptive pattern; and \se is rated according to whether a deceptive pattern is designed through a sequence of activities that mislead a user step by step. 
At last, the impact factor (\imp) is determined by the consequence of a deceptive pattern. At the current stage, we primarily consider three potential adverse consequences that deceptive patterns may bring to users: time wasting, privacy breach, and financial loss. 

After assigning risk levels to each of the sub-factors described above, we determine the factors \adv and \imp through a weighted sum of sub-factors,\footnote{More details on weights, risk values, and scaling factors can be found at \url{https://github.com/GalaxyHBXY/Decoding-Deceptive-Patterns}.} separately. The final risk score $R$ is calculated as follows:
\begin{equation}
R = (\adv - \dr + \alpha) \times (1 + \imp) \times \beta,
\end{equation}
where $\alpha$ adds an offset to $(\adv - \dr)$, keeping its value larger than 0 to ensure that $\imp$ always has a positive impact on the final risk score. $\beta$ is a scaling factor used to normalize the risk score into the interval $[0,10]$ for convenience. Specifically, a risk score of $0 \leq R \leq 3$ reflects a low-risk level, as the adversary's advantage can be mitigated through detection tools (\ie $(\adv - \dr)$ falls in the lower 40 percentile). Conversely, a risk score greater than 7 indicates a high-risk level, as the adversary's advantage can overwhelm the detection tool (making $(\adv - \dr)$ falls in the upper 25 percentile), and the impact of the deceptive pattern is considerable (enlarging $R$ to be higher than 7), either through a more deceptive and misleading design or by leading to a more severe consequence for the user. 

\noindent \textbf{Threats to validity.~}
We note that, at the current stage of our study, we determined the weights and values of each sub-factor through a small-scale consultation involving experts experienced in UI design, software risk assessment, and deceptive pattern detection. We plan to further refine the design of our risk scoring system with a large-scale user study, recruiting participants with diverse knowledge backgrounds and experiences to determine the weights and values more precisely and objectively.

\begin{figure}[t]
\centering
\includegraphics[width=\linewidth]{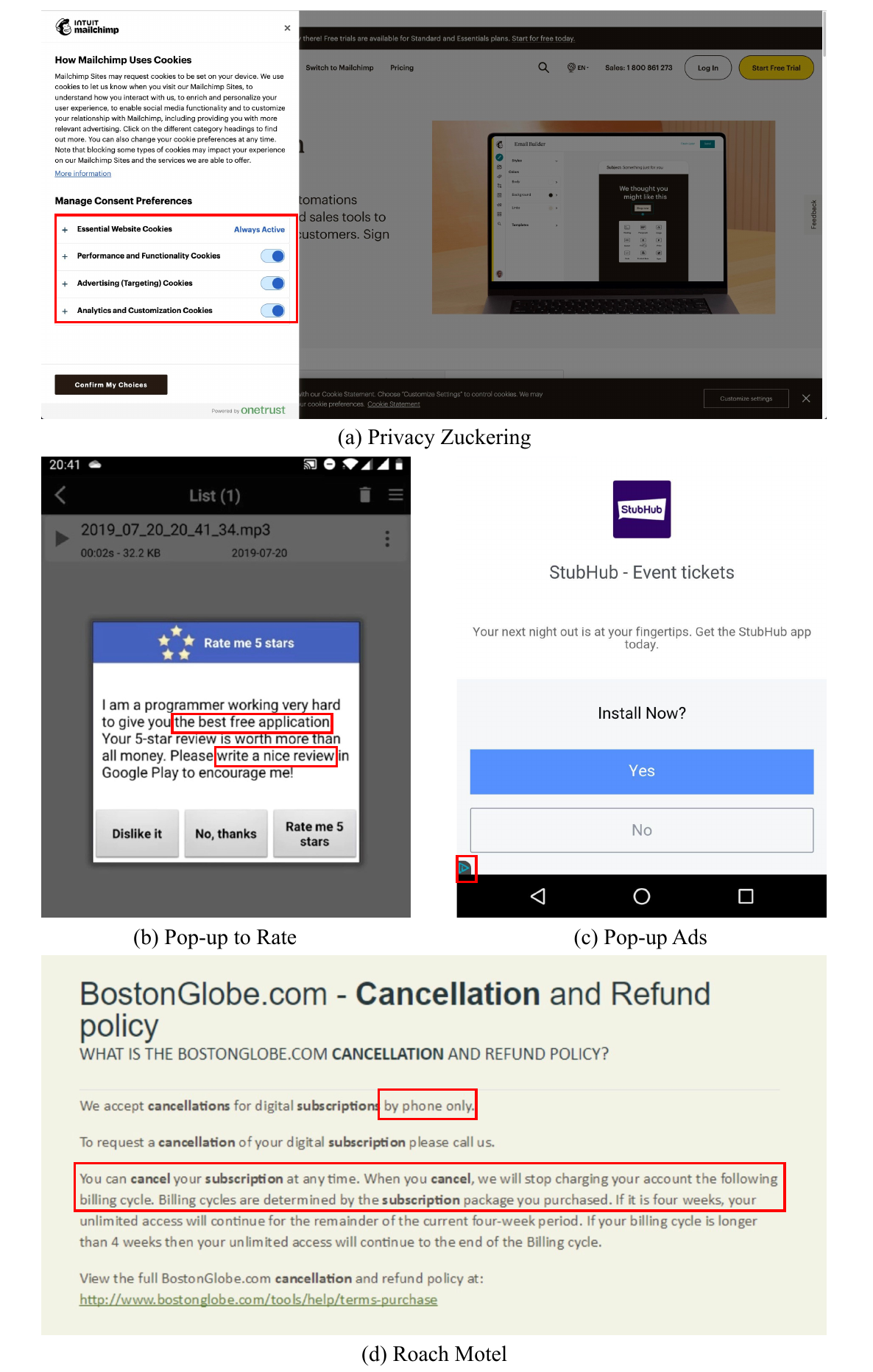}
\caption{Examples of Deceptive Patterns.}
\label{fig_cases}
\end{figure}

\section{Case Studies}
In this section, we apply the proposed deceptive pattern risk scoring system to real world deceptive implementations and present four representative cases to demonstrate the practicability of our scoring system.

\subsection{When both challengers and detectors are effective}

A \textit{Privacy Zuckering} case is demonstrated in Figure~\ref{fig_cases}(a). In this case, a Forced Action deceptive pattern prompts users to undertake certain actions to receive potential benefits.
As illustrated in the UI screenshot, to obtain the basic service of a website, ``Essential Website Cookies'' are forced to be shared, while other cookies are selected to be shared by default. However, after further investigation, we found that most of these ``essential'' cookies are actually non-essential.

As described, with several buttons set to ``On'' by default and a highlighted text ``Always Active'', we consider that the UI has provided information to users to help them recognize that there are several cookies required by the service provider, or at least their information might be collected if they click the bottom banner ``Confirm My Choices''. Therefore, we rate \uf risk as \low and \pk risks as \medium (the user may need some basic privacy knowledge to recognize the trick). The deceptive pattern here is implemented with a static UI, thus we rate the \se risk as \low. These factors result in an \adv score of \advPZ, a quite low advantage score as such deceptive pattern implementation is not hard to be identified by a human user. Meanwhile, as reported by \uiguard, an on-the-shelf detection tool can achieve \fscorePZ F-Score on \textit{Privacy Zuckering} deceptive patterns, which is a relatively high performance, comparing to other more complex patterns. We believe that such a good detection performance may related to the detection of specific UI widges and the descriptions of forced actions as they are typical and easy to identify. The consequence of this deceptive pattern case obviously links to privacy breach as cookies are often used for advertising to track user behavior on websites. 
Taking all aspects into account, the risk score for this \textit{Privacy Zuckering} case is calculated as \PZwithC, which is considered a low-level risk. Without considering the adjustment from human factors, the risk score would rise significantly to \PZwithoutC, reporting nearly a medium-level risk. \textbf{Considering that both challengers and detectors are quite effective in identifying such deceptive patterns, we believe that the adjusted lower risk score is reasonable.}

\subsection{When challengers are more effective}
\textit{Pop-up to Rate}, another widely seen deceptive pattern,  typically involves a user interface element that prompts or forces users to provide a rating or review for a product, service, or application. As shown in Figure~\ref{fig_cases}(b), this design is often manipulative in nature and can influence user perceptions artificially. The pop-up might appear at a strategically timed moment, interrupting the user's experience. It can be designed to be intrusive, demanding attention and potentially causing frustration.

The wording of the pop-up message may be crafted to influence users positively, potentially leading to biased or insincere ratings, \eg the wording ``the best free application'' and ``please write a nice review'' demonstrated in the screenshot. Considering that the purpose of such pop-up message are quite clear, we rate both the \uf and \pk risks as \low. Additionally, this example presents a static sequence thus the \se is also set as \low. The \adv score is then calculated as \advPR. As for \dr, we found that \uiguard has a F-Score of \fscorePR on \textit{Pop-up to Rate} category. In terms of consequences, the primary impact is the waste of users' time. 
With all the factors determined, the risk score is obtained as \PRwithC, indicating a boundary medium-level risk. However, if we only considered detection tools (which are not very effective) in this case, the risk score will rise up to \PRwithoutC, showing a quite significant risk, although still stays in medium risk level. \textbf{We would like to argue that it is important to consider human's capability of deceptive pattern recognition especially when detection tools are not performing well to avoid over-estimation of the risk}.

\subsection{When detectors are more effective}
Our third case is a \textit{Pop-up Ads} deceptive pattern from the Nagging category. Nagging aims to interrupt users when they try to convert their intention to actions. As a prominent subcategory of Nagging, in \textit{Pop-up Ads}, developers embed ads in a software implementation to promote their stakeholders' businesses.

As shown in Figure~\ref{fig_cases}(c), a full-screen pop-up ad interrupts the user's normal activities and tries to mislead their intention to install an advertised app, \texttt{StubHub}, with a ```Yes' to install'' button highlighted. Note that, the YourAdChoices Icon\footnote{\url{https://youradchoices.com/}} (the small triangle icon at the corner, highlighted by a red frame) indicates that the current interface is an advertisement, rather than a desired action or functionality of the app. Considering all these misleading UI contents, we rate the risk level of \uf and \pk as \high. Regarding the \se, since the no pre-operation is needed to reach this deceptive pattern, we rate it as \low, resulting in a \adv value as \advPA. On the other hand, such a deceptive pattern is not quite hard to be detected by a static UI analysis tool. \uiguard~\cite{chen2023unveiling} achieves a \fscorePA F-Score in the \textit{Pop-up Ads} category. We believe this is due to the high accuracy in the identification of the YourAdChoices Icon. As for the \imp, in this case, the primary consequence is time wastage for users, while privacy or financial risks are not directly triggered. With all these factors input to our assessment system, this pop-up ads case is scored as \PAwithC, indicating a medium risk of users being tricked.
Through involving the challenger in the risk assessment, the risk score increases slightly from \PAwithoutC (a low-level risk if \adv is not considered) to \PAwithC (a medium-level risk), indicating that \textbf{although a deceptive pattern is detectable by tools on-the-shelf, a normal user is still threatened to be possibly misled}, calling for more attentions from research and industry communities.

\subsection{When neither challengers nor detectors are effective}
A \textit{Roach Motel} case from a media website is shown in Figure~\ref{fig_cases}(d), which belongs to a sub-category of Obstruction where malicious developers intensively making certain tasks difficult for users to satisfied their stakeholder's interest (\eg obstructing subscription cancellation). 

In this case, we rate the risk level of \uf as \high, considering that there are quite limited texts (\eg ``by phone only'') to guide a user on how to cancel the service and obtain a refund in detail, and the user need to navigate the websites to try to find the phone number. However, the prominently displayed ``cancellation and refund policy'' makes users believe that they can cancel the service at any time, ignoring that the service will actually be canceled in the ``next'' billing cycle. Therefore, we believe that preliminary knowledge or experience is needed to help the user recognize the deceptive pattern, and this deceptive pattern requires a user to go through a multi-step dynamic sequence of actions. Accordingly, the \adv is calculated as \advRM.
On the tool side, neither \uiguard{} nor \aidui supports multi-step dynamic detection. Therefore, we assign the lowest F-Score across categories (\fscoreRM) as its $det$ value. Considering potential consequences, this Roach Motel case may significantly waste users' time and jeopardize users' finances. 
After determining all the aforementioned factors, the overall risk value for this case stands at \RMwithC, a high-level risk. If we disregard the challengers' intervention and assume they will randomly guess, the risk value would decrease to a medium level (scored as \RMwithoutC). Through this case, \textbf{we would like to emphasize that involving challengers in the risk assessment complements the view, as it takes the adversary's advantage against human users into consideration}.

\section{Conclusion and Future Work}
In this study, we introduced a novel threat modeling approach to analyze deceptive patterns within the gaming context, which incorporates the adversary, watchdog, and challenger elements within the game. Furthermore, we have implemented a risk scoring system specifically tailored to deceptive pattern occurrences in this gaming framework. We successfully demonstrated the practicality of our scoring system and underscored the importance of considering human factors in the assessment of deceptive pattern risks, as illustrated by four representative cases.
Looking ahead, our future work aims to enhance and refine our risk scoring system by incorporating additional factors and conducting more extensive large-scale studies. We plan to leverage large language models for data synthesis to obtain more comprehensive results. Additionally, we intend to extend our research by integrating our game-based threat model into the identification of deceptive patterns.

\bibliographystyle{ACM-Reference-Format}
\bibliography{references}


\begin{thebibliography}{12}

%%% the Web address.

\ifx \showCODEN    \undefined \def \showCODEN     #1{\unskip}     \fi
\ifx \showDOI      \undefined \def \showDOI       #1{#1}\fi
\ifx \showISBNx    \undefined \def \showISBNx     #1{\unskip}     \fi
\ifx \showISBNxiii \undefined \def \showISBNxiii  #1{\unskip}     \fi
\ifx \showISSN     \undefined \def \showISSN      #1{\unskip}     \fi
\ifx \showLCCN     \undefined \def \showLCCN      #1{\unskip}     \fi
\ifx \shownote     \undefined \def \shownote      #1{#1}          \fi
\ifx \showarticletitle \undefined \def \showarticletitle #1{#1}   \fi
\ifx \showURL      \undefined \def \showURL       {\relax}        \fi
% invisible to TeX
\providecommand\bibfield[2]{#2}
\providecommand\bibinfo[2]{#2}
\providecommand\natexlab[1]{#1}
\providecommand\showeprint[2][]{arXiv:#2}

\bibitem[Chen et~al\mbox{.}(2023)]%
        {chen2023unveiling}
\bibfield{author}{\bibinfo{person}{Jieshan Chen}, \bibinfo{person}{Jiamou Sun}, \bibinfo{person}{Sidong Feng}, \bibinfo{person}{Zhenchang Xing}, \bibinfo{person}{Qinghua Lu}, \bibinfo{person}{Xiwei Xu}, {and} \bibinfo{person}{Chunyang Chen}.} \bibinfo{year}{2023}\natexlab{}.
\newblock \showarticletitle{Unveiling the Tricks: Automated Detection of Dark Patterns in Mobile Applications}. In \bibinfo{booktitle}{\emph{Proceedings of the 36th Annual ACM Symposium on User Interface Software and Technology}}. \bibinfo{pages}{1--20}.
\newblock


\bibitem[{Consumer Policy Research Centre}(2022)]%
        {cprc2022dark}
\bibfield{author}{\bibinfo{person}{{Consumer Policy Research Centre}}.} \bibinfo{year}{2022}\natexlab{}.
\newblock \bibinfo{title}{Duped by design – Manipulative online design: Dark patterns in Australia}.
\newblock
\newblock


\bibitem[Di~Geronimo et~al\mbox{.}(2020)]%
        {di2020ui}
\bibfield{author}{\bibinfo{person}{Linda Di~Geronimo}, \bibinfo{person}{Larissa Braz}, \bibinfo{person}{Enrico Fregnan}, \bibinfo{person}{Fabio Palomba}, {and} \bibinfo{person}{Alberto Bacchelli}.} \bibinfo{year}{2020}\natexlab{}.
\newblock \showarticletitle{UI dark patterns and where to find them: a study on mobile applications and user perception}. In \bibinfo{booktitle}{\emph{Proceedings of the 2020 CHI conference on human factors in computing systems}}. \bibinfo{pages}{1--14}.
\newblock


\bibitem[{European Parliament}(2022)]%
        {ep2022digital}
\bibfield{author}{\bibinfo{person}{{European Parliament}}.} \bibinfo{year}{2022}\natexlab{}.
\newblock \bibinfo{title}{{Digital Services Act}: Regulating platforms for a safer online space for users}.
\newblock
\newblock


\bibitem[Gole et~al\mbox{.}(2022)]%
        {gole2022new}
\bibfield{author}{\bibinfo{person}{Tim Gole}, \bibinfo{person}{Simon Burns}, \bibinfo{person}{Michael Caplan}, \bibinfo{person}{Melissa Fai}, \bibinfo{person}{Andrew Hii}, \bibinfo{person}{Sheila McGregor}, {and} \bibinfo{person}{Lesley Sutton}.} \bibinfo{year}{2022}\natexlab{}.
\newblock \bibinfo{title}{The new {EU} Dark Pattern laws: Changes to the Digital Services Act}.
\newblock
\newblock


\bibitem[Gray et~al\mbox{.}(2018)]%
        {gray2018dark}
\bibfield{author}{\bibinfo{person}{Colin~M Gray}, \bibinfo{person}{Yubo Kou}, \bibinfo{person}{Bryan Battles}, \bibinfo{person}{Joseph Hoggatt}, {and} \bibinfo{person}{Austin~L Toombs}.} \bibinfo{year}{2018}\natexlab{}.
\newblock \showarticletitle{The dark (patterns) side of UX design}. In \bibinfo{booktitle}{\emph{Proceedings of the 2018 CHI conference on human factors in computing systems}}. \bibinfo{pages}{1--14}.
\newblock


\bibitem[Gray et~al\mbox{.}(2023)]%
        {gray2023towards}
\bibfield{author}{\bibinfo{person}{Colin~M Gray}, \bibinfo{person}{Cristiana Santos}, {and} \bibinfo{person}{Nataliia Bielova}.} \bibinfo{year}{2023}\natexlab{}.
\newblock \showarticletitle{Towards a Preliminary Ontology of Dark Patterns Knowledge}. In \bibinfo{booktitle}{\emph{Extended Abstracts of the 2023 CHI Conference on Human Factors in Computing Systems}}. \bibinfo{pages}{1--9}.
\newblock


\bibitem[Mathur et~al\mbox{.}(2019)]%
        {mathur2019dark}
\bibfield{author}{\bibinfo{person}{Arunesh Mathur}, \bibinfo{person}{Gunes Acar}, \bibinfo{person}{Michael~J Friedman}, \bibinfo{person}{Eli Lucherini}, \bibinfo{person}{Jonathan Mayer}, \bibinfo{person}{Marshini Chetty}, {and} \bibinfo{person}{Arvind Narayanan}.} \bibinfo{year}{2019}\natexlab{}.
\newblock \showarticletitle{Dark patterns at scale: Findings from a crawl of 11K shopping websites}.
\newblock \bibinfo{journal}{\emph{Proceedings of the ACM on Human-Computer Interaction}}  \bibinfo{volume}{3} (\bibinfo{year}{2019}), \bibinfo{pages}{1--32}.
\newblock


\bibitem[Rahman-Jones(2024)]%
        {rahman2024taylor}
\bibfield{author}{\bibinfo{person}{Imran Rahman-Jones}.} \bibinfo{year}{2024}\natexlab{}.
\newblock \bibinfo{title}{Taylor Swift deepfakes spark calls in Congress for new legislation}.
\newblock
\newblock


\bibitem[Russell et~al\mbox{.}(2016)]%
        {russell2016cliptography}
\bibfield{author}{\bibinfo{person}{Alexander Russell}, \bibinfo{person}{Qiang Tang}, \bibinfo{person}{Moti Yung}, {and} \bibinfo{person}{Hong-Sheng Zhou}.} \bibinfo{year}{2016}\natexlab{}.
\newblock \showarticletitle{Cliptography: Clipping the power of {kleptographic} attacks}. In \bibinfo{booktitle}{\emph{Advances in Cryptology--ASIACRYPT 2016}}. \bibinfo{pages}{34--64}.
\newblock


\bibitem[Suciu(2023)]%
        {suciu2023x}
\bibfield{author}{\bibinfo{person}{Peter Suciu}.} \bibinfo{year}{2023}\natexlab{}.
\newblock \bibinfo{title}{X's Misinformation Problem Is Getting Worse}.
\newblock
\newblock


\bibitem[Vincent(2022)]%
        {vincent2022france}
\bibfield{author}{\bibinfo{person}{James Vincent}.} \bibinfo{year}{2022}\natexlab{}.
\newblock \bibinfo{title}{France fines Google and Facebook for pushing tracking cookies on users with dark patterns}.
\newblock
\newblock


\end{thebibliography}

\end{document}